\documentclass[preprint,pra,showpacs,superscriptaddress]{revtex4}
\usepackage{amsmath,amssymb,latexsym,amsfonts}
\usepackage{epsfig}
\begin{document}
\title{Simulation of attosecond streaking of electrons emitted from a tungsten surface}
\date{\today}

\author{C. Lemell}
\affiliation{Institute for Theoretical Physics, Vienna University of 
Technology, Wiedner Hauptstra\ss e 8-10, A--1040 Vienna, Austria}
\author{B. Solleder}
\affiliation{Institute for Theoretical Physics, Vienna University of 
Technology, Wiedner Hauptstra\ss e 8-10, A--1040 Vienna, Austria}
\author{K.\ T\H{o}k\'{e}si}
\affiliation{Institute of Nuclear Research of the Hungarian Academy of 
Sciences, (ATOMKI), H--4001 Debrecen, P.O.Box 51, Hungary}
\author{J. Burgd\"orfer}
\affiliation{Institute for Theoretical Physics, Vienna University of 
Technology, Wiedner Hauptstra\ss e 8-10, A--1040 Vienna, Austria}

\begin{abstract}
First time-resolved photoemission experiments employing attosecond streaking of electrons emitted
by an XUV pump pulse and probed by a few-cycle NIR pulse found a time delay of about 100
attoseconds between photoelectrons from the conduction band and those from the 4f core level of
tungsten. We present a microscopic simulation of the emission time and energy spectra employing
a classical transport theory. Emission spectra and streaking images are well reproduced. Different
contributions to the delayed emission of core electrons are identified: larger emission depth,
slowing down by inelastic scattering processes, and possibly, energy dependent deviations from the
free-electron dispersion. We find delay times near the lower bound of the experimental data.
\end{abstract}

\pacs{79.60.Bm,72.10.-d,78.47.J-}
\maketitle

\section{Introduction}
Photoemission from solid surfaces employing photon energies ranging from the extended
ultraviolet (XUV) to X-ray energies is a well established diagnostics tool to explore the
electronic band structure \cite{ref1}, the crystal structure \cite{ref2}, and its chemical composition \cite{ref3}.
High-resolution angle resolved photoemission can provide detailed information on band gap
and conical intersections, e.g.\ in single-layer graphene \cite{ref4}. EXAFS allows probes of the local
environment by spectral fluctuations due to interfering paths \cite{ref5}. Until recently, virtually
all of these techniques were based on the extraction of spectral information only, i.e.\ on the
energy distribution $P(E_e)$ for a given initial photon energy $E_\gamma$. The use of pulsed sources on
the femtosecond scale has recently given access to dynamical processes, e.g.\ the formation
of image states \cite{ref6}.

The introduction of the attosecond streaking technique \cite{ref7}, originally developed for gasphase
photoionization, to photoemission from solid surfaces has opened up a new perspective:
observation of electronic motion in condensed matter and near surfaces in real time. The
first proof of principle experiment revealed that XUV photoemission from a tungsten surface
features an intriguing time structure with the conduction band electrons coming first while
electrons from core levels, in the present case mostly 4f and 5p, are delayed by about 100
attoseconds ($110\pm 70$ as) \cite{ref8}.

Analyzing the energy-time spectrum and identifying possile sources of the delay poses a
considerable challenge to a microscopic simulation: apart from the intrinsic difficulties in
accounting for the many-body response on a sub-femtosecond scale, the near infrared (NIR)
probe with intensities of $I = 2\cdot 10^{12}$ W/cm$^2$ is sufficiently strong as to actively modify the
electronic response and emission process. Deviations from a field-free emission scenario are
therefore to be expected. We present in the following a simulation of the attosecond streaking
of tungsten employing a classical transport theory (CTT, \cite{ref9}). It allows to account for elastic
and inelastic scattering processes as well as modifications of emission energy and angular
distributions due to the presence of the streaking field. We find, overall, good agreement
with the attosecond-streaking emision spectrum. The time spectrum shows, indeed, delayed
photoemission with a delay time $\tau\approx 40$ as being at the lower bound of the experimental
value. We present a detailed breakdown of different processes taken into account influencing
the time spectrum and discuss possible additional contributions.

\section{Simulation}
Before presenting the key ingredients entering the theoretical treatment, we briefly review
the experiment underlying the scenario of the present simulation \cite{ref8}.

\begin{figure}[ht]
\centerline{\epsfig{file=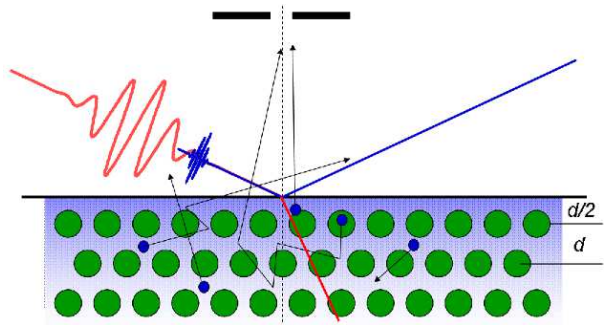,width=8cm}}
\caption{(Color online) Schematic view of experimental setup: A 300 as XUV pulse hits the
surface under a gracing angle of incidence $\theta_{in}\simeq 75.5^\circ$ ionizing target atoms. Excited electrons propagate
in the field of a 10 fs NIR pulse and are detected by a TOF spectrometer mounted perpendicular
to the surface with an acceptance cone of half-width $\Delta\theta=5^\circ$.\label{fig1}}
\end{figure}

Two collinear linearly polarized laser pulses with the polarization direction in the plane
of incidence were directed on a W(110) surface under a grazing angle of incidence (Fig.\
\ref{fig1}).
The two pulses were an XUV pulse at 91 eV with a full width at half maximum (FWHM) of $\sim 6$ eV
and a pulse duration $\tau_{XUV} \simeq 300$ as and a NIR pulse at 702 nm
with $\tau_{NIR} \simeq 10$ fs. The angle of incidence was chosen to be at the Brewster angle for the
probe pulse in order to avoid the influence of the deflected pulse on electrons emitted from
the surface ($\theta_{in} = \arctan (n_{NIR}) \simeq 75.5^\circ$; $n_{NIR} = 3.85$, $k_{NIR} = 2.86$, \cite{ref10};
dielectric function $\varepsilon = \varepsilon_1+i\varepsilon_2 = (n^2-k^2)+2ink)$. This implies
for the XUV pulse ($n_{XUV} = 0.93$, $k_{XUV} = 0.04$;
\cite{ref10}) total reflection at the surface and excitation of photoelectrons by an evanescent wave.
Electrons excited by the pump pulse propagate in the field of the probe pulse modulating
their energy as a function of their emission time (``streaking'', \cite{ref7}). Photoelectrons escaping
the target surface were detected by a time-of-flight spectrometer with the detection direction
normal to the surface. 40 streaking spectra were recorded with relative delays between pump
and probe pulses ranging from $-6$ fs to $+5$ fs with respect to the maxima of the envelope
functions of XUV and NIR pulses.

On top of a strong background signal originating from above-threshold ionization (ATI)
two prominent features were observed in each of the streaking spectra: a lower energy peak
around 55 eV and a higher energy peak at 85 eV. The former was attributed to electrons
excited from 4f core states of tungsten, the latter to electrons from the target conduction
band (consisting of 5d and 6s states).

\begin{figure}[ht]
\centerline{\epsfig{file=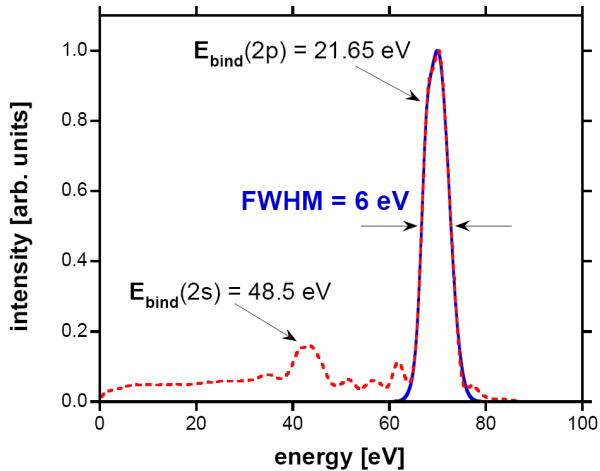,width=8cm}}
\caption{(Color online) Kinetic energy of electrons emitted from Ne irradiated by the XUV pulse
used in the present surface experiment (red dotted line); spectral distribution used in simulation
(blue solid line). From the binding energy of the Ne(2p) level the XUV photon energy of $\sim 91$ eV
was determined.\label{fig2}}
\end{figure}

The incident XUV spectrum has been determined by measuring the kinetic energy of
electrons emitted from a Ne gas target (Fig.\ \ref{fig2}). Two main peaks from emission of 2s and 2p
electrons with binding energies of 48.5 eV and 21.65 eV, respectively, can be distinguished.
We have fitted the peak at 70 eV by a Gaussian function (solid line) from which the XUV
photon energy of about 91 eV and the approximate FWHM of 6 eV were derived.

The simulation involves the solution of the Langevin equation of motion,
\begin{equation}
\dot{\vec p} = -\vec F_{NIR}(\vec r, t) + \vec F_{stoc}(t) + \vec F_{surf}(\vec r)\label{eq1}
\end{equation}
for an ensemble of a large number (typically $10^7$) of initial conditions representing the
primary photoionization electrons. $\vec F_{stoc}(t)$ allows for both elastic and inelastic scattering of
the liberated electron with tungsten cores as well as conduction electrons
\begin{equation}
\vec F_{stoc}(t) =\sum_i \Delta\vec p_i\; \delta(t - t_i)\, ,
\end{equation}
where $\Delta\vec p_i$ is the random collisional momentum transfer as deduced from differential scattering
cross sections. $t_i$ are the random collision times determined from a Poisson distribution
with a mean value given by the mean free path (MFP) or, equivalently, the mean flight
time. The key point of this method is that the stochastic sequence $(\Delta\vec p_i, t_i)$, determined either
from quantum calculations or independent experimental data, allows quantum
scattering information including diffraction to be included in an otherwise classical calculation
(Eq.\ \ref{eq1}, \cite{ref9}). In
between collisions, free electrons with an effective mass (the value of which is discussed below)
propagate within the penetration depth of the time-dependent NIR electric field.
Near the surface, the electrons must overcome the effective surface potential, the asymptotic
part of which is the dynamical image potential \cite{ref11}. This gives rise to the additional
force $\vec F_{surf}(\vec r)$ in Eq.\ \ref{eq1}.

As collisions can result in secondary electron emission, the simulation does not
preserve the number of particles but allows ``en route'' for the generation of additional
trajectories thereby simulating the collision cascade. However, slow secondary electrons
generated outside the energy window of interest are omitted in the following.

\subsection{Primary photoemission as initial condition}
The ensemble of initial conditions represents photoemission of tungsten electrons by a 91
eV XUV pulse which originate from 5 atomic levels: 6s, 5d, 4f, 5p, and 5s. Photoelectrons
from the latter ($E_{bind}(5s) = 75.6$ eV) are submerged in the ATI background and are therefore
omitted in our simulation. Electrons excited from the 4f and 5p levels (referred to in the
following as core levels) give rise to a peak near 55 eV kinetic energy. The binding energies
of the 4f$_{5/2}$ and 4f$_{7/2}$ states are 33.5 eV and 31.4 eV, respectively. The branching ratio
for photon energies around 100 eV of $R_I = I_{7/2}/I_{5/2} = 1.56$ was taken from experimental
photoemission data \cite{ref12}. With a binding energy of 36.8 eV the energy of excited 5p$_{3/2}$
electrons overlaps with the distributions of 4f electrons. Due to the smaller photoexcitation
cross section its relative importance is about a factor 3 smaller than the combined 4f levels
\cite{ref13}. Excitation of the 4f and 5p levels with the XUV pump pulse taking the 4f-branching
ratio into account leads to a slightly asymmetric energy distribution of primary electrons
around 58 eV. For brevity, we refer to this overlapping distribution in the following as the
core level.

\begin{figure}[ht]
\centerline{\epsfig{file=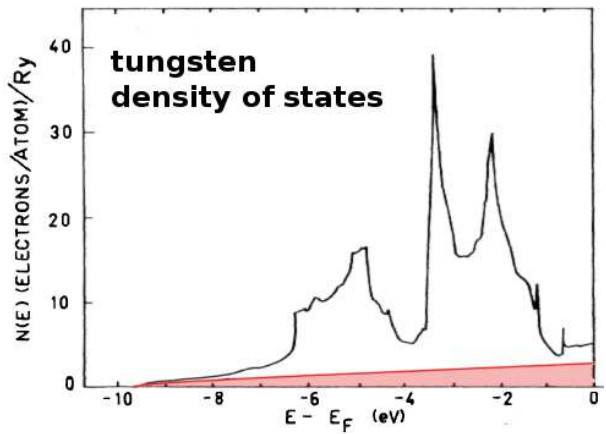,width=8cm}}
\caption{(Color online) Occupied part of the density of states of tungsten \cite{ref14}. States below the red solid line (shaded area) originate from the 6s band, states above the line from the (more localized) 5d level.\label{fig3}}
\end{figure}

The 5d (4 electrons per atom) and 6s (2 electrons per atom) levels of tungsten form the
conduction band. Its density of states (DOS) was taken from \cite{ref14} (Fig.\ \ref{fig3}). The Fermi energy
is 9.75 eV while the workfunction of W(110) is taken to be 5.25 eV \cite{ref15}. We split the DOS in
two components: a free-electron gas like subband containing the 6s electrons (DOS $\propto \sqrt{E}$;
area below red line) and a more localized component representing the 5d electrons. A similar
decomposition of the DOS with a slightly higher value for the Fermi energy has already been
proposed earlier by Mattheiss \cite{ref16}. He suggested the bottom of the 6s band to lie below the
bottom of the 5d band, as shown in Fig.\ \ref{fig3}.
Density-functional theory calculations using the ABINIT package \cite{ref17} with pseudopotentials
for the tungsten cores \cite{ref18} give a Fermi energy of $E_F = 9.6$ eV in good agreement with
\cite{ref14} and \cite{ref16}.

We note that in the spectral region of interest an additional emission channel may contribute:
Auger decay from the conduction band filling a 5s hole created by photoionization
(binding energy of the 5s level $\sim 75.6$ eV). The expected energy range for this Auger emission
would be 45 to 65 eV. We expect, however, that the refilling of 5s holes will proceed
predominantly via the 5s5p$^2$ decay channel. This is also supported by XPS spectra of the
W 4f region in which no signatures of Auger electron emission have been seen (e.g.\ \cite{ref19}).
Therefore, we do not include this channel in our simulation.

\begin{figure}[ht]
\centerline{\epsfig{file=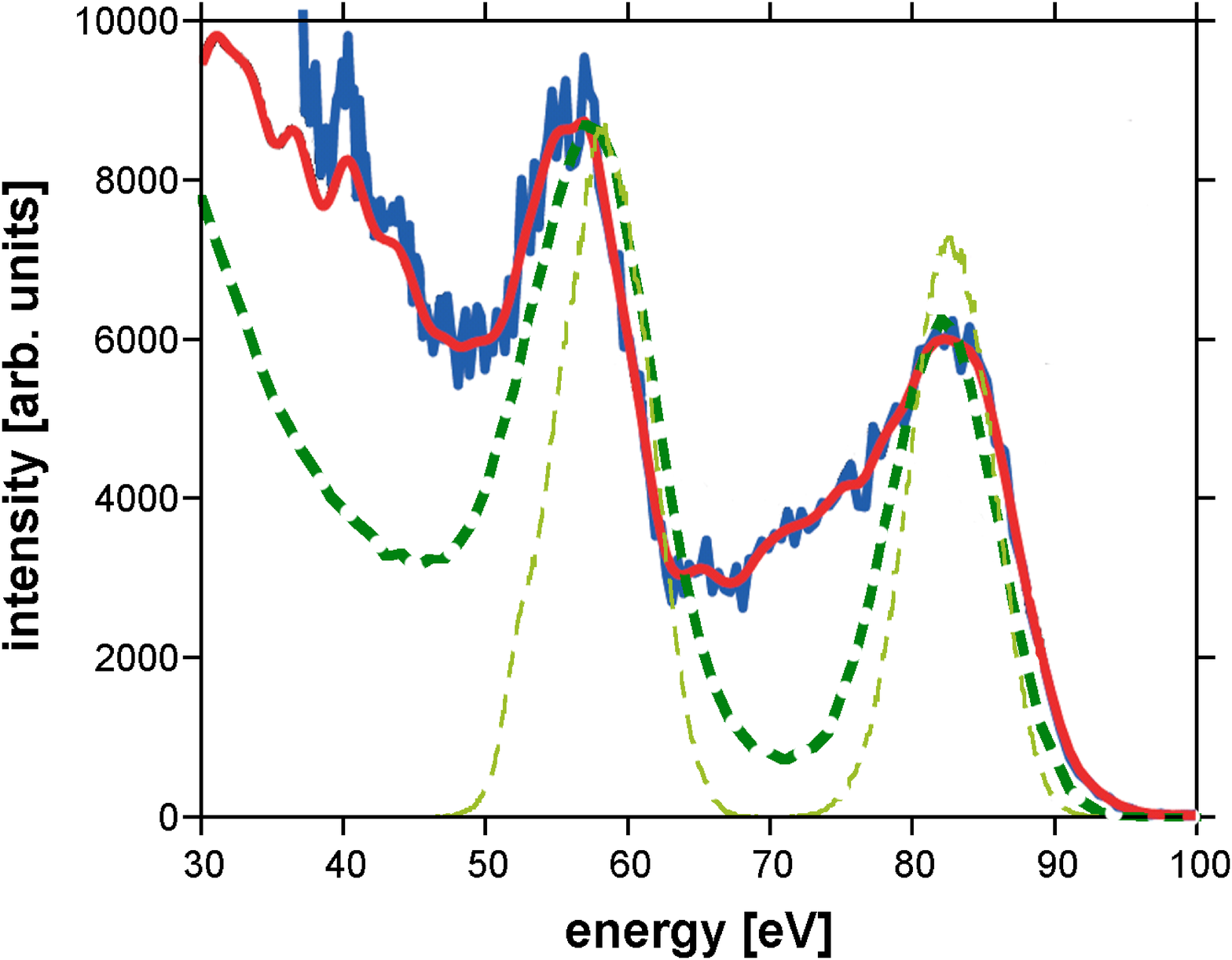,width=8cm}}
\caption{(Color online) Energy spectra of electrons emitted by an XUV laser pulse in absence of
the NIR streaking pulse from a tungsten surface. Experimental data with the ATI background subtracted
are indicated by thick solid
lines (blue -- raw spectrum, red -- smoothed raw spectrum), results of the CTT simulation by the
thick dashed line. The thin dashed line represents the excitation spectrum as represented by the
initial conditions.\label{fig4}}
\end{figure}

We set the emission strength (dipole oscillator strength) for each level to be constant
within the spectral width of the XUV pulse. For the relative intensity ratios between different
levels, we have explored two different options. One uses the calculated photoionization
cross sections for atomic tungsten \cite{ref13}. Accordingly, the photoionization cross section for
the 4f subshell of atomic tungsten is by about a factor two larger than for the 5d level and
by more than an order of magnitude larger than for 6s electrons. Using this input we find a
spectral distribution (in absence of any streaking field) in strong disagreement with experimental
data. Experimental photoemission cross sections from the conduction band appear
to be considerably larger than estimated from atomic cross sections. Alternatively, we have adjusted
the strength of the conduction band peak relative to the core peak in order to
reproduce a photoelectron spectrum after excitation by an XUV pulse (in the absence of the NIR
field -- thick red line in Fig.\ \ref{fig4}).
As the experimental spectrum includes the modification of the primary spectrum due
to multiple scattering during transport until electrons cross the metal-vacuum boundary, a
proper comparison must account for these processes as well (for details see below). Including
transport and broadening of the emission spectrum to simulate the detector resolution
(Gaussian distribution, $\sigma = 2.5$ eV) the simulated photoemission spectrum (thick dashed
line) fits the experimental data (raw data with the ATI background subtracted) rather well.
A notable exception is the low-energy shoulder of the 80 eV conduction band peak the origin
of which is presently not known. We note parenthetically that a very recent measurement shows
a much reduced shoulder \cite{cav_new}.

The angular distribution of primary photoelectrons excited by linearly polarized photon
beams is given in dipole approximation by
\begin{equation}
\frac{d\sigma_{n\ell}(\omega)}{d\Omega} = \frac{\sigma_{n\ell}(\omega)}{4\pi}\left(1+\beta(\omega) P_2(\cos \theta)\right) = \frac{\sigma_{n\ell}(\omega)}{4\pi}\left(1+\frac{\beta(\omega)}{2} (3\cos^2 \theta -1)\right)
\end{equation}
where $\sigma_{n\ell}$ is the total photoionization cross section from the $n\ell$ subshell, $P_2$ is the second
order Legendre polynomial, and $\beta(\omega)$ is the energy-dependent asymmetry parameter. $\beta = 0$
describes isotropic emission while a value of $\beta = 2$ gives a pure cosine distribution. Calculated values for
beta for the considered levels vary from $\beta = 0.7$ (5p) to $\beta = 2$ for s-levels \cite{ref13}. Other sources,
however, suggest different values and experimental data are, to our knowledge, missing. We
have therefore performed simulations for the limiting cases of $\beta = 0$ and $\beta = 2$ but have
found no significant influence of the specific choice on our final results. As an evanescent
wave propagates parallel to the surface, its polarization vector is perpendicular to the surface
($\theta =0$). All results presented in Sec.\ \ref{res} were calculated using $\beta = 2$ (cosine distributions
along surface normal) for all levels.

For the modification of the primary photoionization spectrum by multiple scattering and,
in turn, for analyzing the time structure underlying the emission spectrum, the spatial distribution
of the primary excitation, i.e.\ the depth profile of the source is of crucial importance.
Starting positions of electron trajectories were distributed over the 20 topmost layers of a
W(110) crystal (lattice parameter $a_W = 3.16$ \AA, layer spacing in $\langle 110\rangle$ direction $d = 2.24$ \AA).
For the free-electron like 6s component of the conduction band (Fig.\ \ref{fig3}) we use the jellium
approximation, i.e.\ constant density within the target material starting half a layer spacing
above the topmost atomic layer. For the 5d subband of the conduction band as well as
for the core electrons localized source distributions were chosen from Gaussian distributions
with a FWHM of half a layer distance centered at crystal layer positions in agreement with
our ab-initio calculations.

\begin{figure}[ht]
\centerline{\epsfig{file=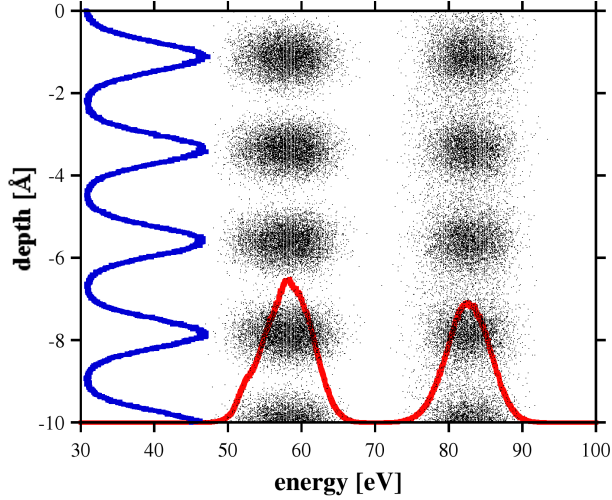,width=8cm}}
\caption{(Color online) Depth-dependent spectral distribution of primary photoelectrons excited
by the XUV pulse with mean energy 91 eV, shown for the five topmost crystal layers. Projections
onto the energy axis and the depth axis indicate the spectral and depth profiles.\label{fig5}}
\end{figure}

The probability of excitation was assumed to be proprtional to the attenuated intensity
of the XUV laser light entering the crystal. For $\theta_{in} = 75.44^\circ$ and $n_{XUV} = 0.93$ total reflection
conditions are fulfilled and only an evanescent wave enters the target. The decay length is
about $\delta\approx 0.55\cdot\lambda\approx 75.6$ \AA\ or, equivalently, more than 30 layer spacings. The depth
dependent reduction in ionization probability has been taken into account, the first five
layers of which are shown in Fig.\ \ref{fig5}.

\subsubsection*{Electron transport}
Electron transport within the target material is calculated using a classical transport
simulation described in detail in \cite{ref20}. In brief, an electron released at its starting point
with initial kinetic energy $E_{kin}$ is subject to elastic and inelastic scattering processes as well
as to deflection of its trajectory due to the influence of the electric field of the probe laser
pulse in the target (Eq.\ \ref{eq1}). The latter is refracted and damped by the target material. For
$n_{NIR} = 3.85$ and $k_{NIR} = 2.86$ at $\lambda = 702$ nm we find a propagation angle of $\theta\approx 14.5^\circ$ with
respect to the surface normal. The NIR intensity after a propagation length $l$ is described
by the Beer-Lambert-Law
\begin{equation}
I(l)=I_0\cdot\exp\left\{-\frac{4\pi k(\lambda)}{\lambda}\cdot l\right\}=I_0\cdot\exp(-\alpha l)
\end{equation}
with the damping constant $\alpha = 0.005$ \AA$^{-1}$ corresponding to a penetration depth of about
85 layers into the target.

Elastic scattering cross sections have been calculated with the ELSEPA package \cite{ref21} using
a muffin-tin potential for the crystal atoms. From the energy-dependent total cross section
and the density of tungsten atoms the elastic mean free path (EMFP) is derived (Fig.\ \ref{fig5}).
The scattering angle in an elastic scattering event is determined by the energy-dependent
differential cross sections.

\begin{figure}[ht]
\centerline{\epsfig{file=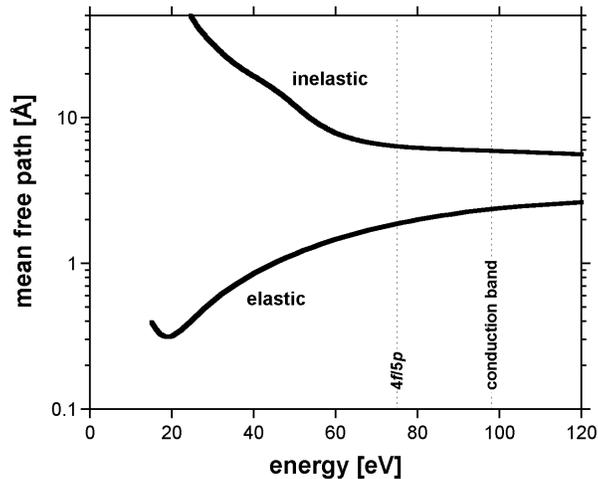,width=8cm}}
\caption{Calculated elastic and inelastic mean free paths of electrons in tungsten. The energy is measured from the bottom of the conduction band ($E_F+W\approx 15$ eV).\label{fig6}}
\end{figure}

The inelastic scattering mean free path (IMFP) is derived from the momentum and energy
dependent dielectric constant of the material $\varepsilon (q,\omega)$ as \cite{ref22}
\begin{equation}
\frac{d^2\lambda^{-1}_{inel}}{dqd\omega}=\frac{1}{\pi Eq}\Im \left\{ -\frac{1}{\varepsilon (q,\omega)}\right\}\Theta[\omega_m(q)-\omega]\label{eq2}
\end{equation}
while the angular distribution of inelastically scattered electrons follows from \cite{ref23}
\begin{equation}
\frac{d\lambda^{-1}_{inel}}{d\Omega}=\frac{1}{\pi^2}\int \frac{d\omega}{q^2}\sqrt{1-\frac{\omega}{E}}\Im \left\{ -\frac{1}{\varepsilon (q,\omega)}\right\} \Theta[E-E_f-\omega]\, .\label{eq3}
\end{equation}
In Eqs.\ \ref{eq2} and \ref{eq3} $E$ is the instantaneous energy of the electron measured relative to the bottom
of the conduction band while $\omega$ and $q$ are the collisional energy and momentum transfers,
respectively. The $\Theta$ (step) functions impose the constraints of energy and momentum conservation
in the scattering event. $\varepsilon (q,\omega)$ is constructed from an extrapolation of the optical
data [$\varepsilon (q=0,\omega)$] for tungsten \cite{ref10} to the $q-\omega$ plane (e.g., \cite{ref24,ref25}).
The resulting inelastic mean free path (Fig.\ \ref{fig6}) in the energy range from 50 to 100 eV is
larger than the EMFP by factors 3 to 10.

Due to the smaller EMFP the average escape depth for electrons emitted
without having suffered energy losses will be smaller than the IMFP. This enhances the
surface sensitivity of the photoemission process. The total mean free path is determined by
\begin{equation}
\frac{1}{\lambda_{tot}}=\frac{1}{\lambda_{el}}+\frac{1}{\lambda_{inel}}
\end{equation}
with $\lambda_{tot}/\lambda_{inel,el}$ being the probability for an (in-)elastic scattering process to happen. Between
subsequent scattering processes electrons propagate in the time-varying electric field
of the NIR laser. Travelling time, energy, and direction of motion are constantly updated.
If an electron reaches the surface of the target it has to overcome the surface barrier which
leads to an additional deflection at the surface. Energy lost in an inelastic scattering event
is transferred to a secondary electron originating at the position of the collision. Its starting
time is given by the time elapsed between start of the primary electron and the scattering
event. As soon as the energy of an electron drops to below 25 eV the trajectory calculation
is stopped.

\begin{figure}[ht]
\centerline{\epsfig{file=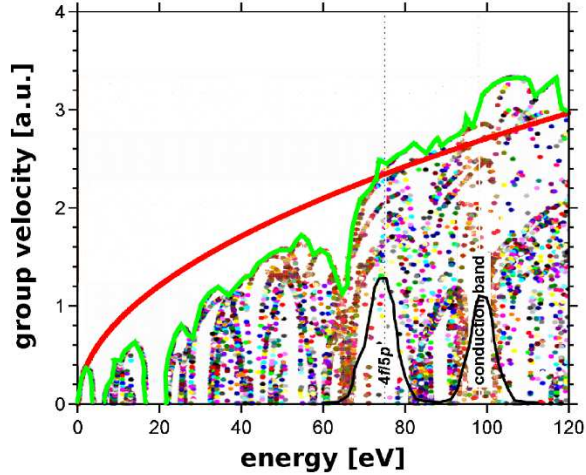,width=8cm}}
\caption{(Color online) Group velocity distributions used in simulation; $v_g = k$ ($m_{eff} = 1$) (red
line), envelope of calculated group velocities (green line) from the dispersion relation of W along
the $\langle 110\rangle$ direction. Energies are measured with respect to the bottom of the conduction band;
$E_F +W \approx 15$ eV with Fermienergy $E_F$ and workfunction $W$.\label{fig7}}
\end{figure}

The effective mass of the energetic electron propagating through the crystal or, equivalently,
its group velocity $v_g$ as determined from the dispersion relation $E(\vec k)$, may have
an important influence on the time dependence of the photoemission. To investigate the
possible influence of the dispersion relation on the observed delay of emitted electrons we
consider two limiting cases: a) a free-particle dispersion relation, $E = k^2/2$, with effective
mass $m_{eff} = 1$ a.u.\ (red line in Fig.\ \ref{fig7}) and b) the distribution of group velocities (green line
in Fig.\ \ref{fig7}) along the $\langle 110\rangle$ direction of W calculated by Silkin et al.\ (supplementary material
to \cite{ref8}).
The latter features dips in the group velocity distribution around 65 eV and 120 eV
with respect to the bottom of the conduction band due to the crystal potential. At both
the expected energies of the core electrons and the conduction band (when measured from
the bottom of the conduction band) we find $v_g$ very close to that of a free electron (Fig.\ \ref{fig7}).
In Ref.\ \cite{ref8}, however, it was assumed that the energy of the core electrons would coincide
with the local minimum of $v_g$ near 65 eV (see Fig.\ \ref{fig7}). Under this assumption larger run
time differences between core electrons and conduction band electrons would result. For test
purposes and in order to maximize the effect of the crystal dispersion relation we followed
this prescription and shifted the envelope of group velocities to match its local minimum
at 65 eV with the energy of the core electrons near 73 eV. The largest run-time differences
reported in Sec.\ \ref{res} were found under these somewhat arbitrary assumptions.

Finally, electrons escaping the surface are subject to the streaking field of the NIR laser
pulse. The latter transfers a momentum of
\begin{equation}
\Delta p=\int_{t_{exc}}^{t_{end}} E(t)\, dt\label{eq8}
\end{equation}
to the electron where the integral is taken from the time of excitation to the conclusion
of the laser pulse. Electrons with a final momentum oriented perpendicular to the surface
with an acceptance angle $\Delta\theta =\pm 5^\circ$ about the surface normal are included in the streaking
image. Parameters describing the NIR pulse (maximum field strength, wavelength, pulse
duration) were taken from \cite{ref8}.

\section{Results}\label{res}

\subsection{Emission energy and emission time}
\begin{figure}[ht]
\centerline{\epsfig{file=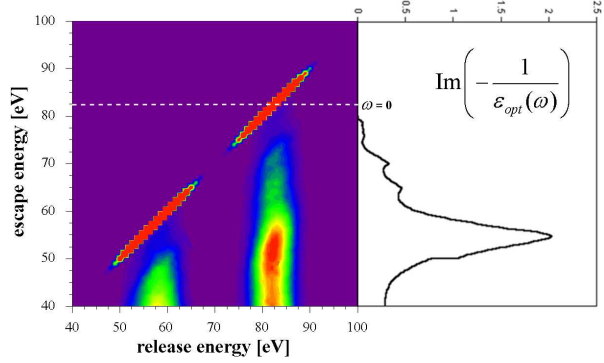,width=8cm}}
\caption{(Color online) Primary excitation energy vs.\ escape energy for photoemission by an XUV
pulse ($\hbar\omega = 91$ eV) from a W surface. The loss function of tungsten (right-hand side) shows a
pronounced excitation (loss channel) at about 25 eV.\label{fig8}}
\end{figure}

We first present results for energy and emission time distributions in the absence of
the streaking field. Energy and emission time spectra are strongly affected by energy loss
properties during transport to the surface. Fig.\ \ref{fig8} depicts the two-dimensional correlated
primary excitation -- final escape energy distribution.
In the absence of inelastic processes
the electron distribution would be located on the diagonal. Due to collisions described by
the energy loss function with a pronounced peak at about 25 eV, a significant portion of
the initial distribution escapes at lower energy. Accidentally, the energy loss peak closely
matches the energy spacing between core and conduction band electrons. Therefore, the
primary conduction band electrons having undergone a single inelastic collision overlap with
the primary energy distribution of the core electrons. The average excitation depth and
traveling time in the target material for such electrons can be larger and will therefore
contribute to the observed time delay. Also the secondary-electron background in the escape
spectrum will be much larger in the spectral region of the core electrons adding to the
observed run-time difference.

\begin{figure}[ht]
\centerline{\epsfig{file=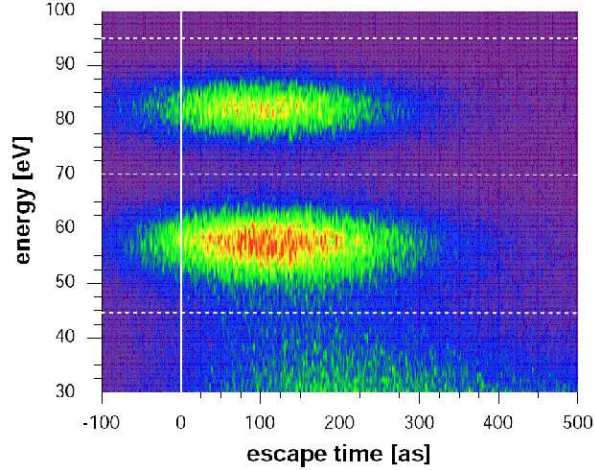,width=8cm}}
\caption{(Color online) Correlated escape-time escape-energy distribution after excitation of a W
surface by an XUV pulse ($\hbar\omega = 91$ eV). $t_{esc} = 0$ corresponds to the peak amplitude of the XUV
pulse.\label{fig9}}
\end{figure}

\begin{figure}[ht]
\centerline{\epsfig{file=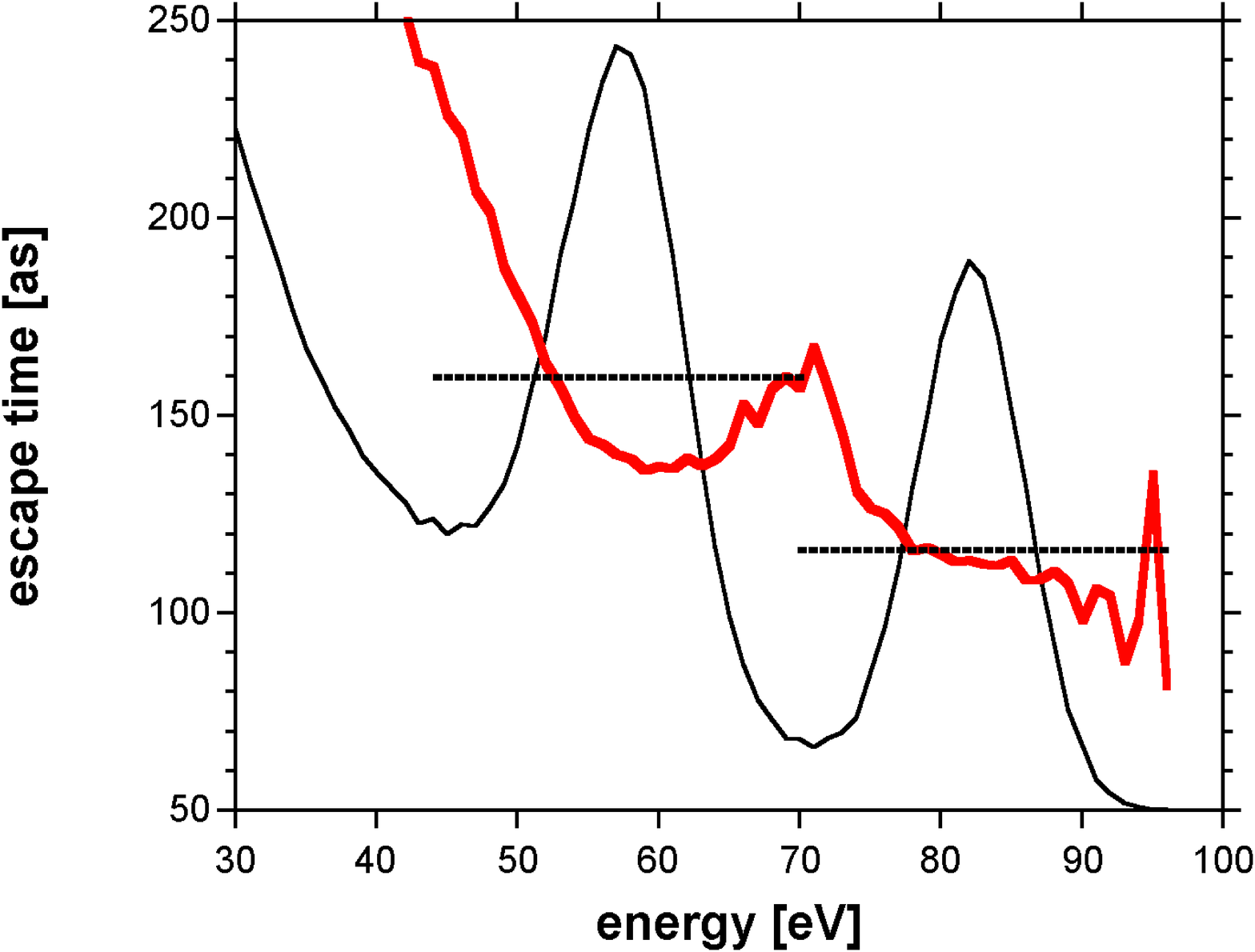,width=8cm}}
\caption{(Color online) Energy spectrum (thin line) and escape times of electrons (thick red line)
excited by an ultrashort XUV pulse. Horizontal lines indicate the escape time averaged over the
core and conduction band peak regions. The mean run-time difference is about 40 as.\label{fig10}}
\end{figure}

The two-dimensional correlated escape-time escape-energy distribution (Fig.\ \ref{fig9}) after the
excitation by the XUV pulse feature consequently a much broader and slightly shifted escape time
distribution of the ``core'' electron peak near 58 eV compared to the conduction-band
distribution near 83 eV.
Note that the $t_{esc} = 0$ line corresponds to the temporal maximum
of the XUV amplitude. The maxima of the distributions at 58 and 83 eV are located close
to 115 and 100 as, respectively. Averaging over the entire peak areas (44--70 eV, 70--95 eV)
results in run times of 157 as and 115 as (see horizontal lines in Fig.\ \ref{fig10}).
The run-time difference ($\sim 42$ as) can be accounted for in part by a simple estimate using the material properties
of W: the average escape depth of electrons is given by their IMFP (see Fig.\ \ref{fig6}). Using the
free-electron dispersion relation the total travel time of electrons with energies of 73 and 98
eV (measured from bottom of conduction band) along their respective IMFP of 6.5 \AA\ and
5.8 \AA\ is about 125 as and 100 as, respectively. This gives a lower bound of the run-time
difference for electrons with escape energies of 58 eV and 83 eV of about 25 as. This runtime
difference is increased by scattering events in the target material which primarily
increase the average run time of electrons ending in the spectral region of the core peak.

Projection onto the energy axis provides the spectrum while temporal information (Fig.\ \ref{fig10})
is extracted by averaging the escape time over the electrons within a given escape energy
bin. While at the maximum of the energy spektrum the run-time difference is only about 20
as, averaging over the entire energy range of the core and conduction band peaks (horizontal
lines) results in a difference of 42 as. This is at the lower bound of the experimental value of
$110 \pm 70$ as. The agreement would be considerably better when using the shifted envelope
of group velocities as discussed above. In this case the run-time difference would be more
than double and would increase to 101 as.

\subsection{Streaking images and center-of-mass motion}
The time structure of the photoelectron emission is extracted in the experiment \cite{ref8} by
attosecond streaking, i.e.\ by time-to-energy mapping in a few cycle NIR pulse. As soon as
photoelectrons are excited they are subject to the NIR streaking field. If a free electron
interacts with the entire laser pulse its momentum remains unchanged. If, however, it is
set free while an electric field amplitude $E(t)$ is present at the position of excitation, a net
momentum is transfered (Eq.\ \ref{eq8}). Varying the delay time τ between XUV pump and NIR probe pulses result in
an oscillation of the final electron energy \cite{ref7}. In the experiment streaking images have been
recorded in delay-time steps of $\Delta\tau = 100$ as. $\tau < 0$ imply the probe pulse to precede the
pump pulse, positive $\tau$ signify the pump pulse preceding the probe.

\begin{figure}[ht]
\centerline{\epsfig{file=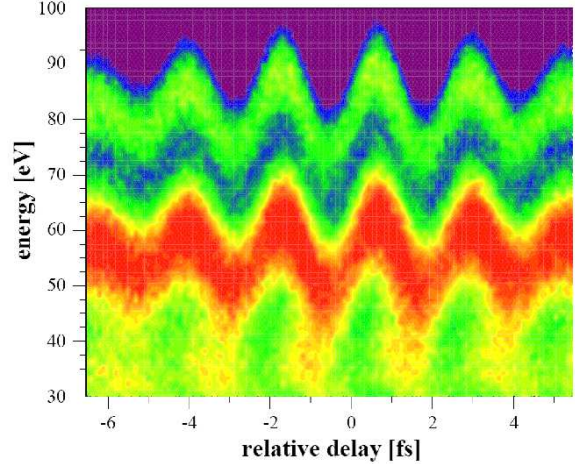,width=8cm}}
\caption{(Color online) Calculated streaking images for electrons emitted by a 300 as XUV pulse
and streaked by a few-cycle NIR pulse. Emission along the surface normal with an acceptance cone
angle of $\theta=0\pm 5^\circ$.\label{fig11}}
\end{figure}

\begin{figure}[ht]
\centerline{\epsfig{file=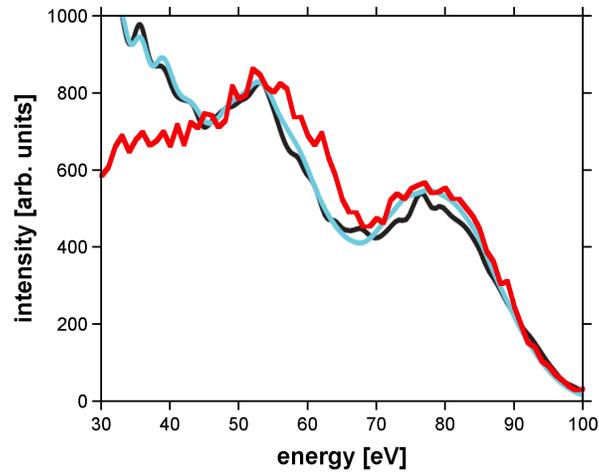,width=8cm}}
\caption{(Color online) XUV pulse induced photoemission spectrum in the presence of a NIR
streaking field. Thick red line: simulated spectrum for electrons emitted in the direction of the
surface normal with an acceptance cone of half-width $\Delta\theta = 5^\circ$; black line: experiment; blue line:
smoothed experimental data.\label{fig12}}
\end{figure}

The characteristic streaking oscillations in the simulated electron energy can be clearly seen (Fig.\ \ref{fig11}).
From the streaking image a mean run-time difference of 33 as between core electrons
(integrated over the energy interval $44 \le E \le 70$ eV) and the conduction electrons ($70 \le
E \le 95$ eV) can be deduced. The values deduced from the streaking image is slightly
lower (by about 10 as) than that directly deduced from the simulated time spectrum. This
discrepancy is not due to the resolution limits of the streaking technique but consequence
of the modification of the electron transport by the streaking field. Electron trajectories are
deflected in the direction of the NIR laser polarization inside the target. As a consequence,
the escape depth of excited electrons is reduced. The point to be emphasized is that the
streaking field not only probes but actively modifies the emission time spectrum. Along the
same lines, the streaking field also alters the final emission energy spectrum (Fig.\ \ref{fig12}).
Taking the streaking field into account, the shape of the electron spectra drastically changes. Almost
perfect agreement between simulation and experiment is achieved. Both the simulation and
the experiment feature broadened core and conduction band peaks (compare Figs.\ \ref{fig5} and
\ref{fig12}). The overall agreement in peak height and width indicates that the present simulation
is capable of accounting for most of the relevant processes governing the XUV-pulse induced
photoemission from a tungsten surface.

\section{Conclusion}
We have presented simulations of the electron excitation, transport, and attosecond
streaking for an XUV-pump-NIR-probe setting near a tungsten surface. This scenario models
the recent experiment by Cavalieri et al.\ \cite{ref8}. We find excellent agreement for the energy
spectrum in the presence of the streaking field. In agreement with the experiment we find
the core electron emission to be delayed relative to the conduction band electrons. Calculated
run-time differences between groups of photoelectrons are at the lower bound of the
error bar of the measurement. Latest experiments indicate a somewhat 
smaller run-time difference of $85 \pm 45$ as \cite{cav_new} reducing the gap between 
measured data and our simulations. Processes responsible for the time delay identified by the
present simulation include the larger emission depth of core electrons, the contribution of
primary emitted conduction electrons slowed down to energies matching emitted core electrons
as well as secondary electrons. When we include a modification of the group velocity
distribution for energies in the region of core electrons as proposed by Silkin et al.\ \cite{ref8},
the delay time increases and is closer to the experimental data. We note, however,
that invoking this correction would require an energy shift of the spectrum for which a
convincing explanation is missing. Possible other mechanisms not yet accounted for include
the influence of the NIR pulse on the primary photoexcitation process and local crystal
field effects on the emission time spectrum. The impact of both effects could be increased by local field
enhancements at the surface (surface plasmons) as recently shown for the nanoplasmonic
field microscope \cite{ref26}. Such fields might influence excitation and transport of photoelectrons
from the conduction band and bound states differently.

\acknowledgments{This work was supported by the Austrian \textit{Fonds zur F\"orderung
der wissenschaftlichen Forschung} under grants no.\ FWF-SFB016 ``ADLIS'' and no.\ 17449,
the European ITS-LEIF network RII3\#026015, and the TeT Grant No. AT-7/2007. One of us
(KT) was also partially supported by the grant ``Bolyai'' from the Hungarian Academy of
Sciences and the Hungarian National Office for Research and Technology. We thank the group of R. 
Kienberger (MPQ Garching) for making their latest data available to us prior to 
publication.}


\begin{thebibliography}{26}
\bibitem{ref1} D.A. Shirley, \prb {\bf 5}, 4709 (1972).
\bibitem{ref2} W. Friedrich, P. Knipping, and M. von Laue, Sitzungsberichte der Mathematisch-Physikalischen Classe der K\"oniglich-Bayerischen Akademie der Wissenschaften zu M\"unchen 1912: 303 (1912).
\bibitem{ref3} K. Siegbahn, \rmp {\bf 54}, 709 (1982).
\bibitem{ref4} T. Ohta, A. Bostwick, J.L. McChesney, T. Seyller, K. Horn, and E. Rotenberg, \prl {\bf 98}, 206802 (2007).
\bibitem{ref5} E.A. Stern, \prb {\bf 10}, 3027 (1974).
\bibitem{ref6} M. Echenique and J.B. Pendry, J. Phys.\ C {\bf 11}, 2065 (1978).
\bibitem{ref7} R. Kienberger, E. Goulielmakis, M. Uiberacker, A. Baltuska, V. Yakovlev, F. Bammer, A. Scrinzi, Th.\ Westerwalbesloh, U. Kleineberg, U. Heinzmann, M. Drescher, and F. Krausz, Nature {\bf 427}, 817 (2004).
\bibitem{ref8} A.L. Cavalieri, N. M\"uller, Th.\ Uphues, V.S. Yakovlev, A. Baltuska, B. Horvath, B. Schmidt, L. Bl\"umel, R. Holzwarth, S. Hendel, M. Drescher, U. Kleineberg, P.M. Echenique, R. Kienberger, F. Krausz, and U. Heinzmann, Nature {\bf 449}, 1029 (2007).
\bibitem{ref9} J. Burgd\"orfer and J. Gibbons, \pra {\bf 42}, 1206 (1990).
\bibitem{ref10} \textit{Handbook of Optical Constants of Solids}, ed.\ E.D. Palik (Academic Press, San Diego, 1985).
\bibitem{ref11} A.G. Eguiluz, \prb {\bf 23}, 1542 (1981).
\bibitem{ref12} H.B. Rose, A. Fanelsa, T. Kinoshita, Ch.\ Roth, F.U. Hillebrecht, and E. Kisker, \prb {\bf 53}, 1630 (1996).
\bibitem{ref13} J.-J. Yeh and I. Lindau, At.\ Data Nucl.\ Data Tables {\bf 32}, 1 (1985).
\bibitem{ref14} E. Christensen and B. Feuerbacher, \prb {\bf 10}, 2349 (1974).
\bibitem{ref15} R.W. Strayer, W. Mackie, L.W. Swanson, Surf.\ Sci.\ {\bf 34}, 225 (1973).
\bibitem{ref16} L.F. Mattheiss, Phys.\ Rev.\ {\bf 139}, A1893 (1965).
\bibitem{ref17} X. Gonze, J.M. Beuken, R. Caracas, F. Detraux, M. Fuchs, G.M. Rignanese, L. Sindic, M. Verstraete, G. Zerah, F. Jollet, M. Torrent, A. Roy, M. Mikami, Ph. Ghosez, J.-Y. Raty, and D.C. Allanet, Comp.\ Mat.\ Science {\bf 25}, 478 (2002).
\bibitem{ref18} A.M. Rappe, K.M. Rabe, E. Kaxiras, and J.D. Joannopoulos, \prb {\bf 41}, 1227(R) (1990); Erratum \prb {\bf 44}, 13175 (1991).
\bibitem{ref19} XPS spectra at \textit{http://www.lasurface.com}.
\bibitem{cav_new} A.L. Cavalieri, R. Ernstorfer, and R. Kienberger, J. Phys.\ B, submitted (2009); private communication (2009).
\bibitem{ref20} B. Solleder, C. Lemell, K. T\"okesi, N. Hatcher, and J. Burgd\"orfer, \prb {\bf 76}, 075115 (2007).
\bibitem{ref21} F. Salvat, A. Jablonski, and C. Powell, Comp.\ Phys.\ Commun.\ {\bf 165}, 157 (2005).
\bibitem{ref22} C.O. Reinhold and J. Burgd\"rfer, \pra {\bf 55}, 450 (1997).
\bibitem{ref23} Z.J. Ding and R. Shimizu, Surf.\ Sci.\ {\bf 222}, 313 (1989).
\bibitem{ref24} C.J. Powell, Surf.\ Sci.\ {\bf 44}, 29 (1974).
\bibitem{ref25} D.R. Penn, \prb {\bf 35}, 482 (1987).
\bibitem{ref26} M.I. Stockman, M.F. Kling, U. Kleineberg, and F. Krausz, Nature Photonics {\bf 1}, 539 (2007).
\end{thebibliography}
\end{document}